# Phase-modulated shaping of narrowband type-I parametric down-converted photons

A T Joseph[1], R Andrews[1], E R Pike[2] and S Sarkar[2]
[1]Department of Physics, The University of the West Indies, St. Augustine, Trinidad and Tobago
[2]Department of Physics, King's College London, Strand, London WC2R 2LS, UK

E-mail: AJoseph@fsa.uwi.tt; randrews@fsa.uwi.tt; roy.pike@kcl.ac.uk; sarben.sarkar@kcl.ac.uk

**Abstract.** We present a general theoretical description of the temporal shaping of narrowband noncollinear type-I down-converted photons using a spectral phase filter with a symmetric phase distribution. By manipulating the spectral phase of the signal or idler photon, we demonstrate control of the correlation time and shape of the two-photon wave function with modulation frequency and modulation depth of the phase distribution.
**PACS codes:** 42.50.Dv; 42.65.Lm

## 1. Introduction

In the process of spontaneous parametric down-conversion [1,2], a pump photon incident on a birefringent crystal is split into two entangled photons (called signal and idler). Entanglement in both wave vector and frequency is due to the conservation of momentum and energy. Type-I down-conversion is characterized by parallel polarizations of the signal and idler photons, whereas in the type-II process, the polarizations are orthogonal. In addition to wave vector and frequency entanglement, noncollinear type-II signal and idler photons are entangled in polarization but in the type-II collinear process, there is no such entanglement [3]. Entangled photons have been used to demonstrate quantum nonlocality [4-7], quantum teleportation [8,9], quantum interference [10], and nonlinear interactions [11].

It is well known that an optical process can be steered towards a particular final state through manipulations of the coherence properties of the optical fields [12,13]. One technique is the modification of the spectral phases of ultrashort pulses [14]. Such pulses have been used to enhance and suppress nonresonant multiphoton transitions [15], to control the excitation of specific rotational Raman transitions [16], to establish constructive and destructive quantum interference between different excitation pathways in an optically dense medium [17], and to investigate dark soliton propagation in optical fibres [18]. Nonlocal pulse shaping with entangled photon pairs has been demonstrated using a monochromator in the path of one of the beams and a piezo-controlled Michelson interferometer in the other beam [19]. The coherence time of the idler photon was measured from the autocorrelation function by varying the spectral resolution of the monochromator. It was found experimentally that the coherence time of the signal beam decreased as the bandwidth of the idler photon was increased. Very recently, the temporal shaping of broadband collinear type-I entangled photons was investigated [20]. The authors demonstrated control of the temporal properties of the entangled photons by using a phase step function of amplitude $\pi$ at the middle of the signal spectrum. They observed a splitting in time of the two-photon wave function into two lobes. In a previous paper [21], we presented a detailed theoretical model of the temporal shaping of spectrally phase-modulated photons produced via the process of collinear type-II SPDC. By manipulating the spectral phase of the signal or idler photons, we showed that it is possible to control the number of photon pairs arriving at a given location with a specific time delay. In this paper, we present a theoretical model that gives a general description of the temporal shaping of narrowband noncollinear type-I down-converted photons using a spectral phase filter with a symmetric phase distribution. We also examine the temporal properties of the phase-modulated entangled photons. In contrast to previous work we demonstrate a new technique of controlling the time separation of the correlated photons. By phase-modulating the idler photons, we show that the time separation of the correlated photons can be controlled by varying the modulation frequency and modulation depth of the spectral filter. We show specifically that it is the modulation frequency which determines which photon of a pair arrives first at the detector and that for large modulation frequencies either signal or idler photons can arrive first.

## 2. Theory

The two-photon correlation function $G^{(2)}(\vec{r}_1, t_1; \vec{r}_2, t_2)$, which describes the coincidence count rate, is proportional to [22]

$$\int_0^{\omega_{k_0}} d\omega_{k_1} \int_0^{\omega_{k_0}} d\omega_{k_2} k_1^2 k_2^2 \exp(-i\omega_{k_1} t_1 - i\omega_{k_2} t_2) \exp(ik_1 r_1 + ik_2 r_2)$$
$$\times \exp\left[-\frac{\varepsilon_\perp^2}{4}(k_1 \sin\theta_1 + k_2 \sin\theta_2)^2\right] \mathrm{sinc}[(k_0 - k_1 \cos\theta_1 - k_2 \cos\theta_2)\varepsilon_3] \quad (1)$$
$$\times \delta(\omega_{k_0} - \omega_{k_1} - \omega_{k_2}).$$

$\vec{r}_1$ and $\vec{r}_2$ are the detector locations, and $t_1$ and $t_2$ are the detection times. The pump beam is taken to be a laser with a transverse Gaussian profile of radius $\varepsilon_\perp / \sqrt{2}$ and photons are generated from a crystal of length $2\varepsilon_3$. The pump beam is monochromatic with frequency $\omega_{k_0}$ and wave vector $\vec{k}_0$. The crystal is embedded in a passive medium whose linear refractive index is the same as that of the crystal [23]. $\vec{k}_1$ and $\vec{k}_2$ are the wave vectors of the signal and idler photons, which emerge from the crystal at angles $\theta_1$ and $\theta_2$; $\omega_{k_1}$ and $\omega_{k_2}$ are the frequencies of the signal and idler photons, respectively.

The idler photon is phase-modulated using a spectral phase filter with a symmetric spectral phase distribution of the form

$$\theta(\omega_{k_2}) = \alpha \cos(\beta \omega_{k_2}), \quad (2)$$

where $\alpha$ is the modulation depth and $\beta$, the modulation frequency (figure 1). For a phase-modulated idler photon, the two-photon correlation function, to a good approximation, is calculated to be proportional to

$$\int dv \exp\left[-i\frac{\omega_{k_0}}{2}(t_1 + t_2) + iv\tau\right] \exp\left[i\left(k_1^* r_1 + k_2^* r_2 + \frac{v}{u} R\right)\right] \exp\left(-\frac{\varepsilon_\perp^2 v^2}{u^2} \sin^2\theta\right)$$
$$\times \exp\left[i\alpha \cos\left(\beta\frac{\omega_{k_0}}{2} - \beta v\right)\right], \quad (3)$$

where $\tau$ is the time delay between the signal and idler photons ($\tau = t_2 - t_1$) and $R = r_1 - r_2$. $u$ is the group velocity of either the signal or idler photon which we assume to be equal, $\vec{k}_1^*$ and $\vec{k}_2^*$ are the phase-matched wave vectors and $v = \omega_{k_1} - (\omega_{k_0}/2)$. We have assumed that the photons detected emerge from the crystal at equal angles, $\theta$, to the pump beam, i.e. the photons are degenerate.

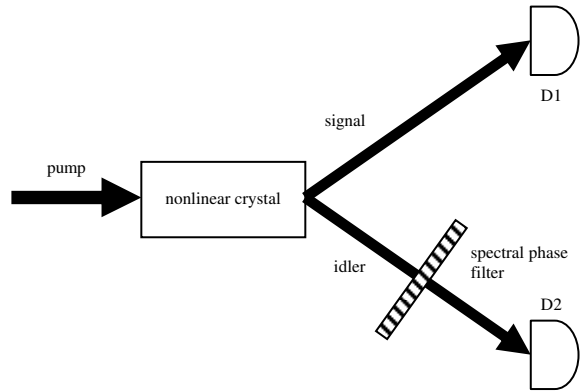

Figure 1 Schematic set-up of an experiment in which a pump beam incident on a nonlinear crystal is down-converted into signal photons and idler photons (in the path of which is a spectral phase filter) that are detected in coincidence at detectors D1 and D2.

After integrating over all $\nu$, (3) becomes proportional to

$$\exp\left(-\frac{u^2\tau^2}{4\varepsilon_\perp^2 \sin^2\theta_1}\right) \sum_{m=-\infty}^{\infty} i^m J_m(\alpha)\exp\left(im\beta\frac{\omega_{k_0}}{2} + \frac{2u^2\tau m\beta - u^2 m^2\beta^2}{4\varepsilon_\perp^2 \sin^2\theta_1}\right), \qquad (4)$$

where, for convenience, we have chosen $r_1 = r_2$. $J_m(\alpha)$ is a Bessel function of order $m$. The absolute square of (4) is proportional to the two-photon coincidence count rate.

## 3. Results

Figure 2 shows the variation of $\tau_{max}$ with modulation depth, $\beta$, where $\tau_{max}$ is the relative time separation between signal and idler photons, at maximum coincidence count rate. In our simulation, $\alpha = 2$, the pump photon is of wavelength 350 nm, $u = 2 \times 10^8$ ms$^{-1}$, $\varepsilon_\perp = 100$ μm and $\theta_1$ is given a value of 15°. We observe an oscillatory dependence of $\tau_{max}$ with $\beta$ in therange 48 fs $\leq \beta \leq$ 53 fs. This shows that it is possible to control the time separation of photon pairs $\tau_{max}$ by varying the modulation frequency within a specified range. It is interesting to note that photon pairs with $\tau_{max} = 0$, i.e. coincident photons, can be obtained for particular values of $\beta$. This would have potential applications in two-photon absorption studies which require photons to arrive at atoms or molecules with controllable time separations.

Figure 3 shows the variation of the count rate with $\tau$ (signal-idler time delay) for modulation depth $\alpha = 0$, $\alpha = 2$ and $\alpha = 10$ for two modulation frequencies $\beta = 50$ fs (figure 3(a)) and $\beta = 53$ fs (figure 3(b)). As expected, for no phase modulation ($\alpha = 0$), the position for maximum two-photon count rate corresponds to $\tau_{max} = 0$. With the use of the spectral phase filter where $\alpha$ is nonzero, different values of $\tau_{max}$ are obtained. This demonstrates again the possibility of controlling $\tau_{max}$ by varying $\alpha$ for a given value of $\beta$. Figure 3 confirms that for any non-zero $\alpha$ when $\beta = 50$ fs, $\tau_{max}$ is negative and it is positive when $\beta = 53$ fs. It is therefore the modulation frequency that determines which photon of a pair arrives first at the detector.

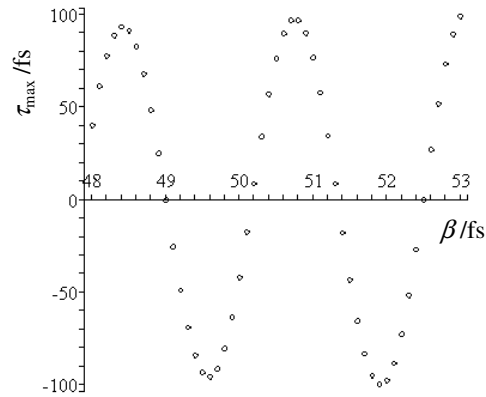

Figure 2 Signal-idler time separation, $\tau_{max}$ against modulation frequency $\beta$

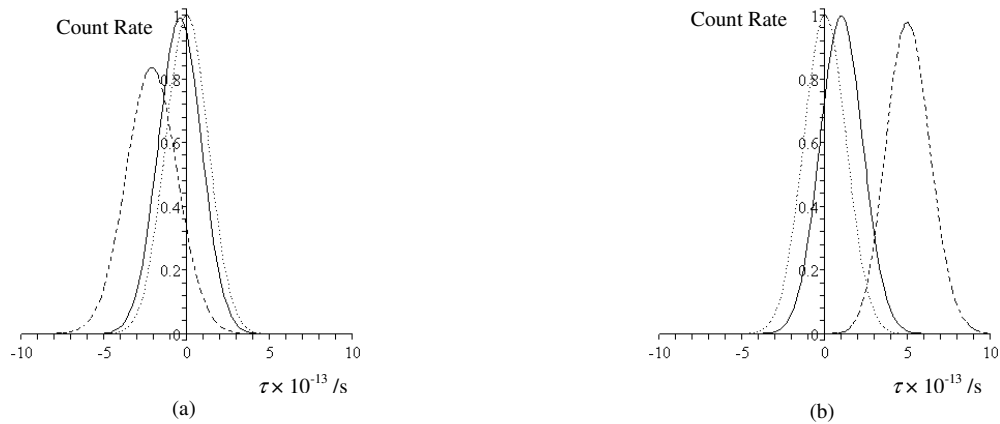

Figure 3 Variation of count rate against signal-idler time delay $\tau$ with
(a) $\beta = 50$ fs for $\alpha = 0$ (dotted line), $\alpha = 2$ (solid line) and $\alpha = 10$ (dashed line) and
(b) $\beta = 53$ fs for $\alpha = 0$ (dotted line), $\alpha = 2$ (solid line) and $\alpha = 10$ (dashed line).

Illustrated in figure 4 are plots of the count rate versus time delay for modulation frequencies of $\beta = 50$ fs, $\beta = 300$ fs and $\beta = 1000$ fs and $\alpha = 2$. It is found that for $\beta > 200$ fs, there is a splitting in time of the two-photon correlation function into two or more lobes. This splitting in time indicates the arrival of signal photons earlier or later than idler photons.

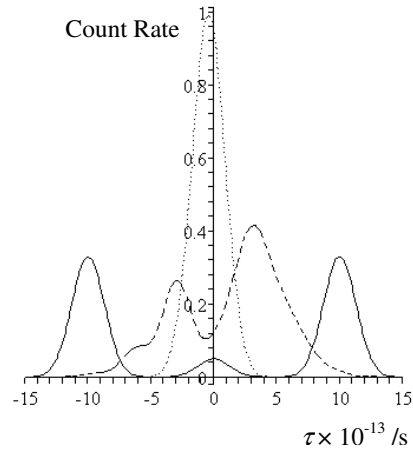

Figure 4 Variation of count rate with time delay for $\beta = 50$ fs (dotted line), $\beta = 300$ fs (dashed line) and $\beta = 1000$ fs (solid line). $\alpha = 2$.

## 4. Conclusions

We have given a general theoretical framework for describing the temporal shaping of narrowband noncollinear photons produced via type-I parametric down-conversion. To achieve such control, a spectral phase filter with symmetric phase distribution and variable modulation depth and frequency was placed in the path of the idler photon. We have shown that the temporal shape of the two-photon wave function depends on the modulation frequency and depth. For a fixed modulation depth of $\alpha = 2$, the time separation can be controlled in the range $-105$ fs $\leq \tau_{max} \leq 105$ fs by varying $\beta$ in the range 48 fs $\leq \beta \leq$ 53 fs. It is therefore possible to obtain a relative time separation that is negative, zero or positive by varying the modulation frequency. We believe that this technique for varying the time separation can be used as an alternative method for examining fourth-order interference effects. Additionally, we have shown that for a given value of the modulation frequency $\beta$, $\alpha$ can be used to vary $\tau_{max}$ over positive values or negative values depending on the choice of $\beta$. We have also demonstrated that for large values of $\beta$, it is possible for the signal or idler photon to arrive earlier for a given value of $\beta$. Since these possibilities are indistinguishable, interference effects can be explored.